\documentclass[12pt,prb,
 amsmath,amssymb,
 aps,
]{revtex4-2}

\usepackage{gensymb}
\usepackage{textcomp}
\usepackage{textgreek}
\usepackage{graphicx}
\usepackage{epstopdf}
\usepackage{dcolumn}
\usepackage{bm}
\usepackage[utf8]{inputenc}
\usepackage{newunicodechar}
\usepackage{xcolor}
\usepackage{url}
\usepackage{soul}
\usepackage{placeins}
\usepackage{cancel}
\usepackage[normalem]{ulem}
\usepackage{siunitx}

\begin{document}

\title{Visualization of Tunable Electronic Structure of Monolayer TaIrTe$_4$}

\author{Sandy Adhitia Ekahana$^{1,2,*}$}
\author{Aalok Tiwari$^{1,*}$}
\author{Souvik Sasmal$^{1,*}$}
\author{Zefeng Cai$^{3}$}
\author{Ravi Kumar Bandapelli$^{1}$}
\author{I-Hsuan Kao$^{1}$}
\author{Jian Tang$^{7}$}
\author{Chenbo Min$^{3}$}
\author{Tiema Qian$^{6}$}
\author{Kenji Watanabe$^{4}$}
\author{Takashi Taniguchi$^{5}$}
\author{Ni Ni$^{6}$}
\author{Qiong Ma$^{7,8}$}
\author{Chris Jozwiak$^{2}$}
\author{Eli Rotenberg$^{2}$}
\author{Aaron Bostwick$^{2}$}
\author{Simranjeet Singh$^{1}$}
\author{Noa Marom$^{1,3,9}$}
\author{Jyoti Katoch$^{1,\dagger}$}

\affiliation{$^{1}$Department of Physics, Carnegie Mellon University, Pittsburgh, Pennsylvania 15213, USA}
\affiliation{$^{2}$Advanced Light Source, E. O. Lawrence Berkeley National Laboratory, Berkeley, California 94720, USA}
\affiliation{$^{3}$Department of Materials Science and Engineering, Carnegie Mellon University, Pittsburgh, Pennsylvania 15213, USA}
\affiliation{$^{4}$Research Center for Electronic and Optical Materials, National Institute for Materials Science, 1-1 Namiki, Tsukuba 305-0044, Japan}
\affiliation{$^{5}$Research Center for Materials Nanoarchitectonics, National Institute for Materials Science,  1-1 Namiki, Tsukuba 305-0044, Japan}
\affiliation{$^{6}$Department of Physics and Astronomy and California NanoSystems Institute, University of California Los Angeles, Los Angeles, CA, USA}
\affiliation{$^{7}$Department of Physics, Boston College, Chestnut Hill, MA, USA}

\affiliation{$^{8}$The Schiller Institute for Integrated Science and Society, Boston College, Chestnut Hill, MA, USA}
\affiliation{$^{9}$Department of Chemistry, Carnegie Mellon University, Pittsburgh, Pennsylvania 15213, USA}

\thanks{$^*$These authors contributed equally to this work.}
\thanks{$^\dagger$Corresponding author: jkatoch@andrew.cmu.edu}

\date{\today}

\begin{abstract}
\clearpage
Monolayer TaIrTe$_4$ is an ideal platform to explore the tunability of band topology where charge doping converts the quantum spin Hall insulator (QSHI) state into a dual QSHI state. However, experimental visualization of its band structure has been lacking despite its importance in interpreting this topological phase transition.
Here we report the band structure of monolayer TaIrTe$_4$ using micro-focused angle-resolved photoemission spectroscopy ($\mu$ARPES). Observed dispersions agree with the Heyd–Scuseria–Ernzerhof (HSE) calculations, establishing the insulating ground state without strong electronic correlations. Furthermore, we show an electron–hole asymmetry in the doping response, where hole doping is readily induced by electrostatic gating, whereas electron doping via gating and  alkali-metal deposition yields band renormalization instead of a rigid shift. Fractional-charge calculations demonstrate that added electrons shrink the band gap and drive band renormalization. Thus, our work identifies the microscopic mechanism by which induced charges reshape the band topology of monolayer TaIrTe$_4$, beyond the rigid shift behavior.

\end{abstract}

\maketitle

\section*{Introduction}
Theoretical and experimental confirmations of topological materials have largely relied on the non-interacting one-particle picture \cite{hasan2010colloquium, qi2011topological, armitage2018weyl}, which typically treats electron correlations as perturbative effects on the non-interacting ground state \cite{raghu2008topological,pesin2010mott,neupert2022charge}. This one-electron picture also remains applicable under various sample modifications such as elemental doping \cite{hsieh2009a,chen2010massive}, surface modification \cite{valla2012photoemission}, strain \cite{yan2015prediction}, and application of electric \cite{checkelsky2009quantum,steinberg2010surface, brahlek2015transport, ong2021experimental}, or magnetic fields \cite{huang2023angle}. Meanwhile, explicitly incorporating strong electron correlations into the theoretical calculations has led to exotic correlated states such as fractional topological insulator \cite{levin2009fractional,neupert2011fractional} and novel phenomena in twisted bilayer van der Waals (vdW) systems \cite{rafi2011moire,  xie2021fractional, cai2023signatures}, which lie beyond the scope of the non-interacting picture.

Previously, bulk TaIrTe$_4$ and NbIrTe$_4$ emerged as type-II Weyl semimetal candidates, predicted to undergo transformation into a quantum spin Hall insulator (QSHI) in their monolayer forms, based on non-interacting one-particle calculations \cite{koepernik2016tairte,liu2017van}. Their experimental realization \cite{koepernik2016tairte,khim2016magnetotransport,liu2017van,haubold2017experimental,belopolski2017signatures,ma2019nonlinear,li2017ternary,zhou2019nonsaturating,ekahana2020topological,zhang2022colossal} has largely confirmed the reliability of the non-interacting bulk band structure calculation. For example, the surface Fermi arcs, characteristic of pairs of oppositely charged bulk Weyl points, have been reported in this framework \cite{haubold2017experimental,ekahana2020topological} and experimentally probed in bulk crystals using angle-resolved photoemission spectroscopy (ARPES) and its time-resolved variant, to access unoccupied conduction band states \cite{belopolski2017signatures}. Recent transport studies on monolayer TaIrTe$_4$, a ternary transition metal chalcogenide (TTMC)\cite{liu2017van}, have  revealed a correlated topological phase emerging upon electron doping \cite{tang2024dual}.
This phase, absent in non-interacting band structure calculations, was attributed to a charge instability arising when the van Hove singularity in the conduction band becomes occupied, leading to a 15-unit-cell charge order \cite{tang2024dual}. Combined theoretical and experimental efforts further uncovered a rich phase diagram in monolayer TaIrTe$_4$, encompassing quantum spin Hall insulator (QSHI), higher-order topological insulator (HOTI), trivial insulator, and metallic phase, tunable via applied strain, electron interactions, and carrier doping \cite{li2025interaction}. The experimental observation of a QSHI state, its dual counterpart, and the ability to manipulate the competing quantum phases, have revived the interest in the interplay between correlations and topology as a function of layer number, twist angle, and strain, despite challenges arising from its air sensitivity and the difficulty of isolating the monolayer. To date, the direct experimental visualization of the electronic band structure of monolayer TaIrTe$_4$ is still missing, to confirm the charge neutral state and explore its behavior under different carrier densities (hole or electron) using electrostatic gating and alkali metal deposition.

 \section*{Results and Discussion}
Here we visualize the electronic structure of monolayer TaIrTe$_4$ in a functional field effect device, using in operando $\mu$ARPES with high spatial resolution of 2 $\mu$m. The observed band structures at both the charge-neutral and hole-doped states closely resemble the one-particle picture obtained with density functional theory (DFT) using the Heyd-Scuseria-Ernzerhof (HSE) hybrid functional. This reaffirms the reliability of the one-particle picture framework when electron exchange effects are included in the HSE functional. Upon electron-doping via gating, we find that instead of filling the conduction band, the existing valence bands are renormalized.

\subsection*{In-operando $\mu$ARPES of monolayer TaIrTe$_4$ field effect device}
\begin{figure*}
\includegraphics[width=0.95\textwidth]{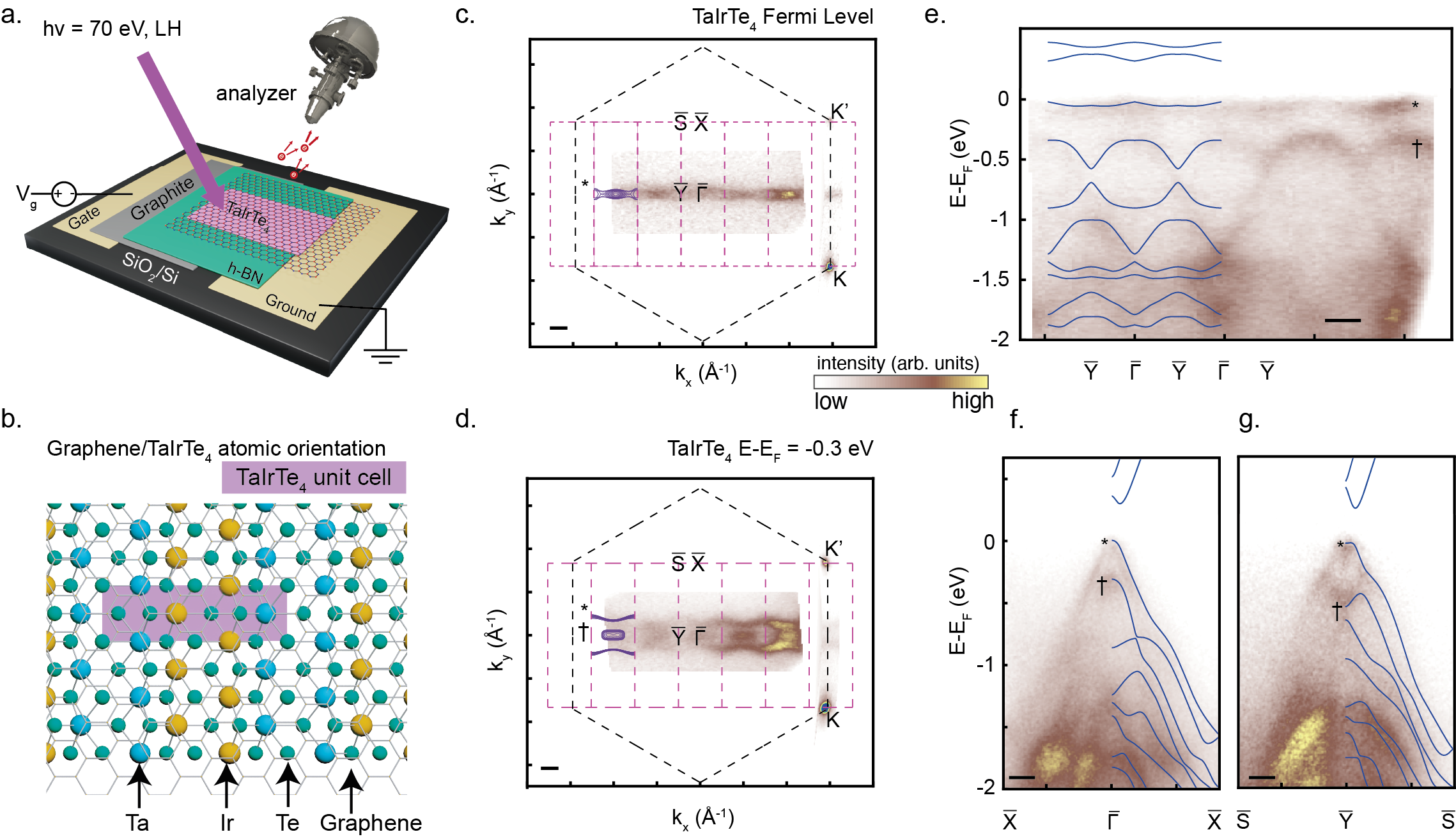}
\caption{\label{figure1}\textbf{Overview of the in operando $\mu$ARPES setup of a monolayer TaIrTe$_4$ field effect device}. a. Top view schematic of TaIrTe$_4$ device integrated with ARPES measurements. b. Crystallographic orientation of the monolayer TaIrTe$_4$ relative to graphene layer in the measured device. c. Fermi surface of the TaIrTe$_4$ with its Brillouin zone and the graphene Brillouin zone shown. d. Isoenergy surface of the TaIrTe$_4$ at $\mathrm{-0.3~eV}$. Asterisk marks indicate the calculated bandstructure from DFT calculation with integration energy window of $\mathrm{0.05 ~eV}$ (purple lines). e. TaIrTe$_4$ ARPES dispersion along the $\mathrm{\bar{Y}\bar{\Gamma}}$ line, f. the $\mathrm{\bar{\Gamma}\bar{X}}$ line, and g. the $\mathrm{\bar{Y}\bar{S}}$ line with the respective calculated bandstructures (blue lines) showing good agreement. The momentum scale-bar is $0.2\,\text{\AA}^{-1}$.}
\end{figure*}

The schematic of the measured monolayer TaIrTe$_4$ device and the ARPES setup are shown in Fig.\ref{figure1}a. As ARPES is a surface sensitive technique, the air-sensitive monolayer TaIrTe$_4$ was protected after cleaving in the glovebox, by capping it with a single layer of graphene sheet (see Fig. \ref{sfigure4} for further reference). The surface quality of the sample was further maintained by minimizing its exposure to air and annealing the device (at $\mathrm{200\degree C}$) prior to the measurements (See Methods \ref{devicefab} for details of heterostructure and device fabrication). The ARPES measurements were performed using a photon energy of $\mathrm{70~eV}$ with linear horizontal (LH) polarization (photon energy dependent measurement at the $\bar{\Gamma}$ point is shown in Fig. \ref{S1a}a). The analyzer was positioned normal to the sample surface when probing the TaIrTe$_4$ bands and aligned at an angle when accessing the K point of the graphene, without moving the beam spot on the sample. An optical image of the monolayer flake and the measured device is shown in Fig.\ref{sfigure4}a,d, where the circle indicates the spatial position on the device where all the ARPES data acquisition was carried out. The colored outlines indicate the individual layers of the stack from top to bottom: graphene (green), TaIrTe$_4$ (red), hBN (sky blue), and graphite (black). The device is back-gated through the graphite electrode, while the graphene is grounded. The hBN layer, with a thickness between 8 and 10 nm, serves both as the dielectric layer and as an atomically flat substrate for TaIrTe$_4$ (see Fig.\ref{figure1}a and Fig.\ref{S1a}d).

\subsection*{Electronic ground state of graphene encapsulated monolayer TaIrTe$_4$ }
The presence of TaIrTe$_4$ is confirmed by the core level spectrum shown in Fig.\ref{sfigure4}f, which exhibits distinct peaks corresponding to all three constituent elements. Fig.\ref{figure1}b illustrates the crystallographic orientation of the monolayer TaIrTe$_4$ relative to the graphene top layer. The pink rectangle represents the conventional unit cell of the monolayer TaIrTe$_4$, drawn to scale with the graphene lattice. The reciprocal space orientations and the corresponding Brillouin zones (BZs) of the TaIrTe$_4$ and graphene are displayed in Fig.\ref{figure1}c (at the Fermi level, $E_F$) and Fig.\ref{figure1}d (at an isoenergy of $\mathrm{0.3~eV}$ below the Fermi level). It is immediately apparent that the system is not commensurate. The simulated isoenergy contours, integrated over an energy window of $\Delta E=0.05\mathrm{eV}$, for the monolayer TaIrTe$_4$ at $E_F$ and at $E-E_F = -0.3 \mathrm{eV}$ are also shown. The simulation is in good agreement with the experimental data. These overview maps highlight the quasi-one-dimensional electronic structure of the TaIrTe$_4$, where multiple BZs of TaIrTe$_4$ are contained inside the first BZ of graphene. 

Fig.\ref{figure1}e shows the dispersive bands extending across several BZs along the short reciprocal lattice constant direction,  $\mathrm{\bar{Y}}\bar{\Gamma}\mathrm{\bar{Y}}$, in good agreement with the DFT calculations (blue lines). The apparent missing-band periodicity in the ARPES data is attributed to the photoemission matrix element effects. The slow-electron-velocity dispersive directions, $\mathrm{\bar{X}\bar{\Gamma}\bar{X}}$ and $\mathrm{\bar{S}\bar{Y}\bar{S}}$ are shown in Fig.\ref{figure1}f and g, respectively, together with the corresponding computed monolayer bandstructures, which show excellent consistency with experiment. The hBN signal is generally absent when probing regions of the stack where graphene and TaIrTe$_4$ are positioned on top (see Fig.\ref{sfigure4}e for probing location). The bands marked with * and $\dagger$ in Fig.\ref{figure1}f,g, exhibit quasi one-dimensional (1D) characteristics near their band tip. Both correspond to two hole-like pockets modulated along the $\mathrm{\bar{Y}}\bar{\Gamma}\mathrm{\bar{Y}}$ direction, the fast dispersive direction. In particular, the *-band hosts the valence band van Hove singularity (vHS), as confirmed by the calculated density of states (Supplementary Fig.\ref{sfigure2}a), located approximately $0.05 \mathrm{eV}$ below the Fermi level, though not resolved within the experimental energy resolution. 
 \subsection*{DFT Calculations Overview}
It is worth noting that our DFT band structure, which is calculated using the HSE hybrid functional, shows better agreement with the ARPES data than obtained using the Perdew-Burke-Ernzerhof (PBE) semi-local functional (see Fig.\ref{S1b} and Fig.\ref{sfigure3}). Previous studies of monolayer, bilayer, and bulk TaIrTe$_4$ have primarily employed PBE or the local density approximation (LDA) \cite{koepernik2016tairte, kumar2021room, li2025interaction, jiang2025probing, Shipunov2021JPCL, guo2020quantum, zhao2022tairte4} (an orbital-decomposed HSE density of states (DOS) was presented in the supporting information (SI) of Ref. \cite{zhao2022tairte4} albeit without spin-orbit coupling). It is well established that semi-local exchange-correlation functionals tend to underestimate band gaps due to the self-interaction error (SIE)\cite{yu2020machine}. For monolayer TaIrTe$_4$, our PBE calculations produce an indirect gap of 0.0461\,eV, whereas our HSE calculations predict a larger gap of 0.237\,eV. The HSE result indicates that the charge-neutral monolayer TaIrTe$_4$ is gapped, consistent with the QSHI scenario. We further evaluated the impact of spin-orbit coupling (SOC) on the monolayer TaIrTe$_4$ band structure. We find that SOC is the determining factor driving the system from a semimetal to a QSHI, consistent with results reported by Guo et al \cite{guo2020quantum}. However, previous studies may have overlooked SOC and its consequences. For example, in the PBE band structure reported by Zhao et al. \cite{zhao2022tairte4}, the omission of SOC results in a semimetal characterized by band crossings along the $\Gamma$–X and Y–S(R) paths. Therefore, we included SOC in all of our PBE and HSE calculations to capture the correct band topology.
\subsection*{Electron–hole asymmetry under electrostatic gating}

\begin{figure*}
\includegraphics[width=0.6\linewidth]{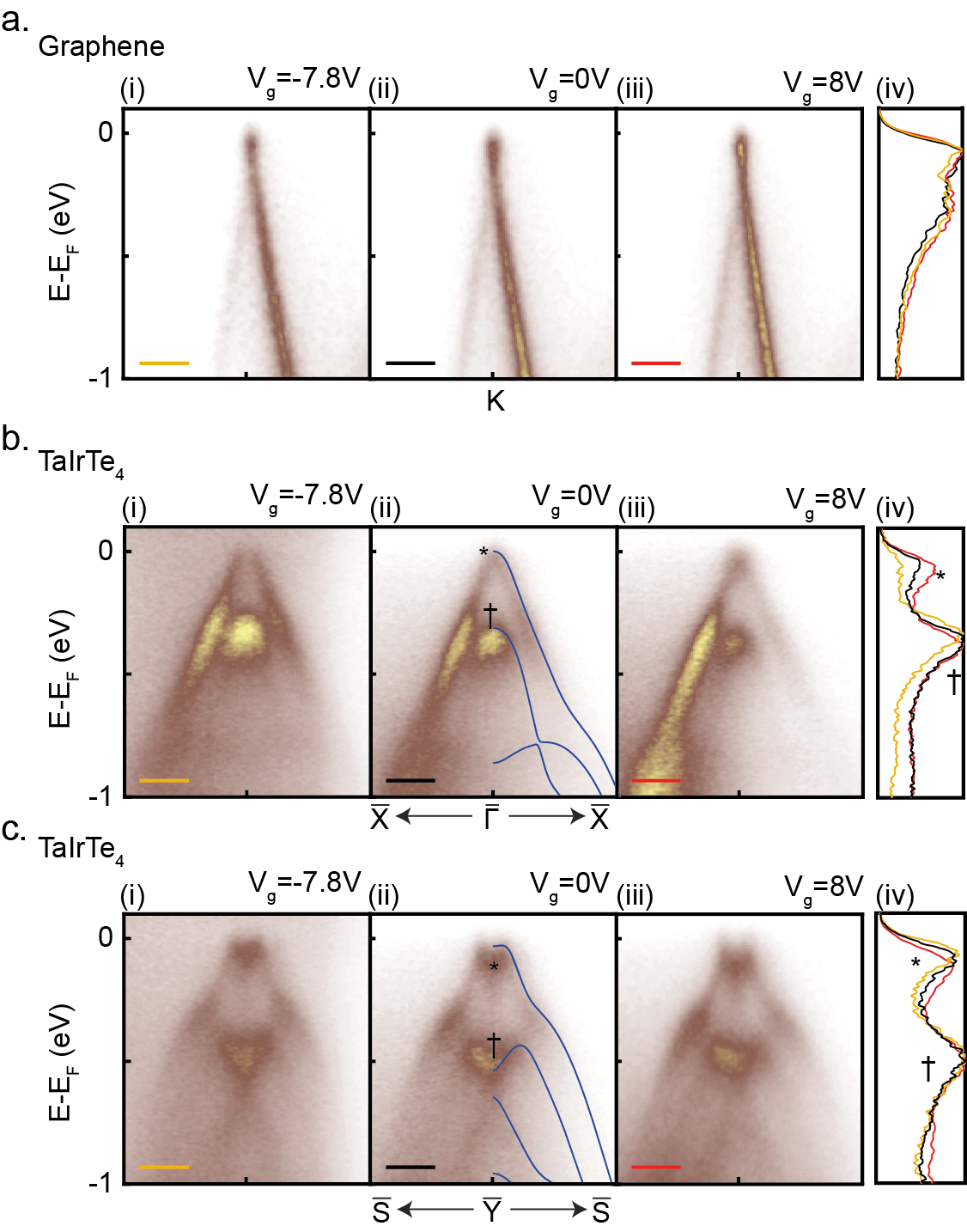}
\caption{\label{figure2}\textbf{Evolution of the gate-dependent band structure of graphene and monolayer TaIrTe$_4$} Gate-dependent bandstructure of a. Graphene, b. TaIrTe$_4$ $\mathrm{\bar{X}\bar{\Gamma}\bar{X}}$ cut, and c. TaIrTe$_4$ $\mathrm{\bar{S}\bar{Y}\bar{S}}$ cut with (i) as $-7.8\mathrm{V}$ (yellow momentum scale-bar), (ii) as zero bias $\mathrm{0V}$ (black momentum scale-bar), and (iii) as $\mathrm{8V}$ (orange momentum scale-bar). (iv) The normalized energy distribution curve (EDC) from the middle momentum (K for graphene and $\mathrm{\bar{\Gamma}}$ for the TaIrTe$_4$) shown with different color according to the momentum scale-bar line color. The symbols * and $\dagger$ indicate the two hole pocket bands. In general, upon increasing the gate voltage from negative to positive, the band shifts from the hole-doped to the electron-doped regime. The momentum scale-bar is $0.2\text{\AA}^{-1}$.}
\end{figure*}

Fig. \ref{figure2} summarizes the ARPES band structure  of graphene and TaIrTe$_4$ along the high symmetry directions and their evolution under applied electric field gating via the graphite back gate. The corresponding results are shown in Fig.\ref{figure2}a-c. Fig.\ref{figure2}a shows the graphene band dispersion around the $\mathrm{K}$ point while Fig.\ref{figure2}b and Fig.\ref{figure2}c present the TaIrTe$_4$ dispersions along the $\mathrm{\bar{X}\bar{\Gamma}\bar{X}}$ and $\mathrm{\bar{S}\bar{Y}\bar{S}}$ directions, respectively. The applied voltages are $V_g$= -7.8V for (i) with yellow momentum scale-bar, neutral for (ii) with black momentum scale-bar, and $V_g$= 8V for (iii) with orange momentum scale-bar. The normalized energy distribution curves from the middle momentum (K for graphene and $\mathrm{\bar{\Gamma}}$ for TaIrTe$_4$) are shown in panel (iv) where the three different applied voltages are shown according to the color. It should be noted that under zero applied gate voltage, the graphene is already slightly electron doped. In general, both graphene and TaIrTe$_4$ display the expected response to electrostatic gating i.e., electrons are depleted (accumulated) under negative (positive) gating. However, the magnitude of the Fermi level shift, particularly for the graphene bands, is smaller than that observed in comparable graphene-based devices subjected to similar back gate voltage values \cite{nguyen2019visualizing,muzzio2020momentum,dale2022correlation} (see Supplementary Fig.\ref{S1c} for an estimation of the Fermi level shift). Specifically, although the thinner hBN dielectric compared to Ref. \cite{muzzio2020momentum} should induce a higher charge density, this expected increase is not observed in the graphene ($\Delta n \approx 1.4\times 10^{12} / \mathrm{cm}^{2}$ as mentioned in Supplementary Fig.\ref{S1c} compared to $\Delta n \approx 4\times 10^{12} /\mathrm{cm}^{2}$ in Ref. \cite{muzzio2020momentum}). This suggests that there is charge transfer in between the graphene and the monolayer TaIrTe$_4$.

The response of the monolayer TaIrTe$_4$ along the $\mathrm{\bar{X}\bar{\Gamma}\bar{X}}$ direction when negative voltage ($V_g$= -7.8V) is applied, indicates that electron depletion (or hole doping) occurs, as the *-band and the $\dagger$-bands shift closer to Fermi level in Fig.\ref{figure2}b(i) compared to the zero voltage ($V_g$= 0V) condition in Fig.\ref{figure2}b(ii). The EDC in panel (iv) also shows that the $\dagger$-band peak lies at lower binding energy while the *-band loses its intensity agreeing with the dispersion cut of panel (i). The energy shift is approximately $0.04 \mathrm{eV}$, corresponding to a removal of $n_e \approx 0.1\times10^{14} /\mathrm{cm^2}$ from the monolayer TaIrTe$_4$ (see Fig.\ref{sfigure2}b), while the overall band dispersion still closely resembles the HSE prediction. This carrier density is higher than what is visibly depleted from the graphene layer, since depleting $n_e \approx 0.1\times10^{14} /\mathrm{cm^2}$ in graphene would require a much larger energy shift (about $0.3 \mathrm{eV}$) than what is observed. This confirms that the graphene and monolayer TaIrTe$_4$ share the induced charges and act as the top metallic plate of the capacitor, with the hBN as the dielectric. This can explain the smaller electron depletion on graphene as it also depletes the monolayer TaIrTe$_4$.

Under positive gate voltage ($V_g$= 8V), the * and $\dagger$-bands of TaIrTe$_4$ exhibit a different response. The $\dagger$-band's energy position remains largely unchanged, as supported by the EDC peak position and width, but its spectral intensity decreases relative to the V-shaped *-band ``legs". This suggests that a positive gate voltage (or electron doping) modifies the matrix elements, thereby altering the relative intensity, which needs further theoretical exploration to understand it. Additionally, the tip of the *-band maximum appears broadened along energy direction and increases in intensity, which arises from the additional charges induced by the gating (symmetric with the hole-doped situation). Since our neutral HSE calculation predicts a band gap (Fig.\ref{figure1}f-g) for monolayer TaIrTe$_4$, this suggests that the additional charges induced in the monolayer TaIrTe$_4$ renormalize the valence band (alter the band shape) before visibly filling the unoccupied conduction band.

The gate voltage dependence of the TaIrTe$_4$ band dispersion along the $\mathrm{\bar{S}\bar{Y}\bar{S}}$ direction is shown in Fig.\ref{figure2}c, revealing trends consistent with those observed in Fig.\ref{figure2}b. Upon applying negative gate voltage ($V_g$= -7.8V), the M-shaped *-band loses intensity and appears flatter, resembling the calculated *-band in Fig.\ref{figure2}c(ii). The EDC also shows that the yellow line *-band peak is slightly closer to the Fermi position than the black line *-band peak. When positive gate voltage $V_g$= 8V is applied, the M-shaped *-band becomes more pronounced, with its tip sharpening into a two-horn-like structure, as supported by the red line EDC that it shifts towards higher binding energy. Meanwhile, the $\dagger$-band remains at approximately the same binding energy for both negative and positive applied voltage. This behavior supports the interpretation that electron doping and also hole doping renormalize the monolayer TaIrTe$_4$ bands and do not necessarily raises its Fermi level, as expected from a rigid band shift of one-particle picture.

\subsection*{High density electron dosing via cesium deposition}
\begin{figure*}
\includegraphics[width=0.95\textwidth]
{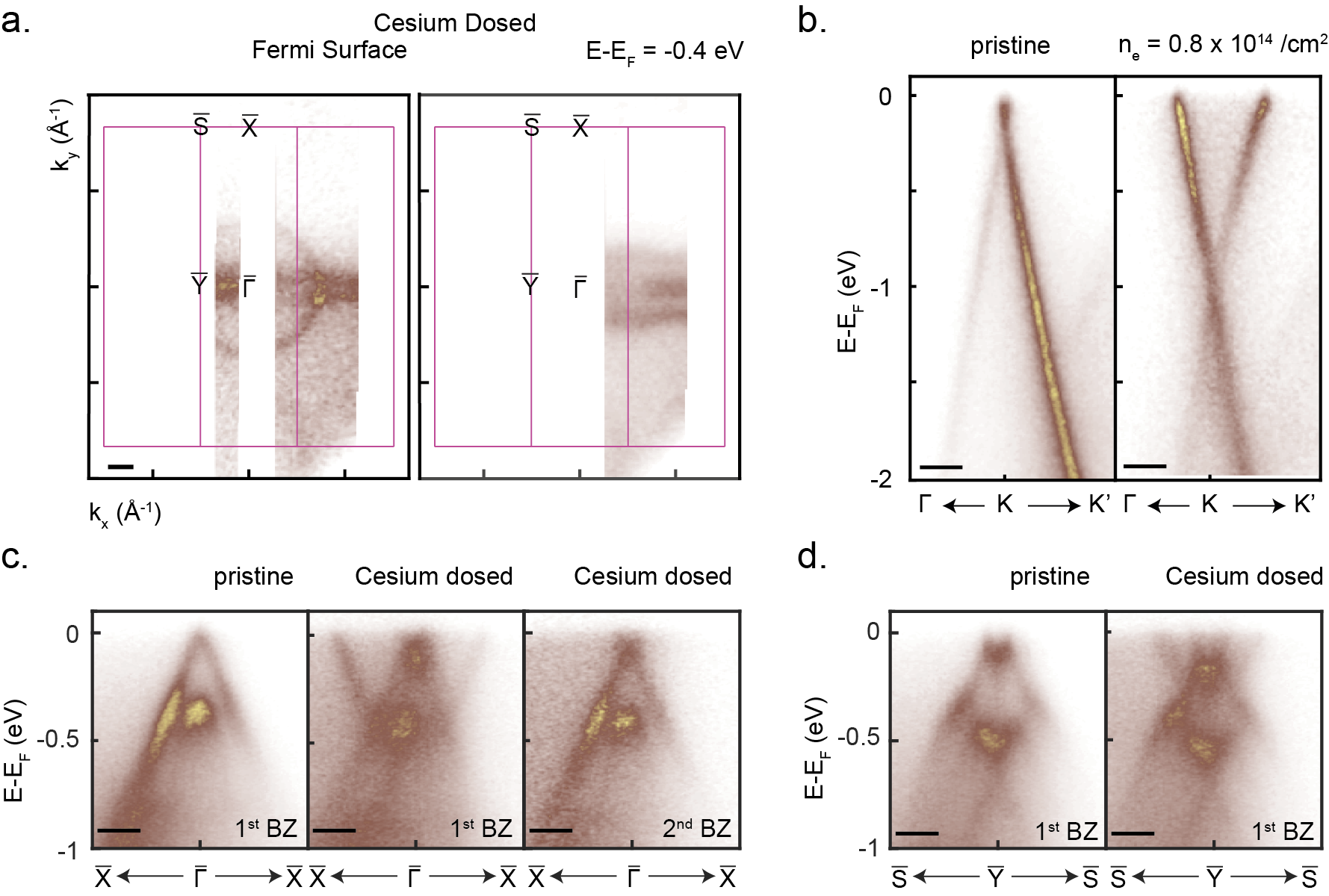}
\caption{\label{figure3}\textbf{Cesium dosed bandstructure of graphene and TaIrTe$_4$ revealing a cesium induced electron pocket around the $\mathrm{\Gamma}$ point.} a. Fermi level after cesium dosing revealing the Fermi surface of the cesium induced electron pocket decorating the pristine TaIrTe$_4$ band, and the isoenergy of the cesium dosed system at $\mathrm{-0.4~eV}$ from the new Fermi level. Cesium dosed bandstructure of b. graphene, c. TaIrTe$_4$ $\mathrm{\bar{X}\bar{\Gamma}\bar{X}}$ cut from the first Brillouin zone and second Brillouin zone, d. TaIrTe$_4$ $\mathrm{\bar{S}\bar{Y}\bar{S}}$ cut. The graphene's Fermi level rises significantly as compared to the Fermi level of the TaIrTe$_4$. The momentum scale-bar is $0.2\,\text{\AA}^{-1}$}
\end{figure*}

Next, in an attempt to further raise the Fermi level of the monolayer TaIrTe$_4$, cesium (Cs) atoms were deposited on the surface of the same TaIrTe$_4$ device, and the evolution of the electronic structure of graphene and TaIrTe$_4$ was examined under in-situ sequential electron doping via alkali metal deposition. First, the presence of Cs atoms on the top surface is confirmed by the core-level spectra shown in Fig.\ref{S1a}d. While the alkali atoms are distributed randomly over the top graphene layer, each Cs atom is expected to donate an extra electron to the graphene/TaIrTe$_4$ heterostructure, which would increase the overall Fermi level of the system. However, the ARPES measurements shown in Fig.\ref{figure3}a reveal that the Fermi surface of TaIrTe$_4$ becomes decorated with a new electron pocket, whose size does not correspond to the reciprocal lattice of TaIrTe$_4$. Meanwhile, the isoenergy surface at $E-E_F=-0.4 \mathrm{eV}$ closely reproduces that in Fig.\ref{figure1}d at $E-E_F=-0.3 \mathrm{eV}$, indicating a small shift of approximately $0.1 \mathrm{eV}$ towards higher binding energy for monolayer TaIrTe$_4$. In contrast, a much larger energy shift of about $0.8\mathrm{eV}$ is observed in the graphene Fermi level, as shown in Fig.\ref{figure3}b, corresponding to an electron transfer of approximately $n_e = 0.8\times 10^{14} \mathrm{/cm^2}$ to the top graphene layer. 

Fig.\ref{figure3}c presents the electron-doped TaIrTe$_4$ band dispersions obtained through Cs deposition along the $\mathrm{\bar{X}\bar{\Gamma}\bar{X}}$ directions in different BZs, while Fig.\ref{figure3}d shows the corresponding cuts along the $\mathrm{\bar{Y}\bar{S}\bar{Y}}$ direction in the first BZ. The Cs-dosed spectra exhibit an overall energy shift of approximately $0.1 \mathrm{eV}$, consistent with the iso-energy surface. We note that the additional electron pocket appearing between $\mathrm{\bar{X}}$ and $\bar{\Gamma}$ in Fig. \ref{figure3}c is observed only in the first BZ and is absent in the next neighboring BZ. The electron pocket has the size of 
$k_F = 0.36 \pm 0.01~\text{\AA}^{-1}$ which implies that the pocket has electron density of $n_e \approx 2.10\times10^{14} /\mathrm{cm^{2}}$. Beyond the $0.1 \mathrm{eV}$ shift, an enhanced spectral intensity is observed near the tip of the *-band which is not repeated in the $2^{\mathrm{nd}}$ BZs suggesting a non TaIrTe$_4$ origin. The tip shape in the $2^{\mathrm{nd}}$ BZ Fig. \ref{figure3}c resembles the *-band modification upon applying positive gate voltage shown in Fig.\ref{figure2}b(iii) indicating that some of the electrons from the Cs are donated to the monolayer TaIrTe$_4$. Furthermore, the tip now weakly suggests a ``U" shape indicating a conduction band is approaching the valence band and starting to become occupied. The $\mathrm{\bar{Y}\bar{S}\bar{Y}}$ cut with electron doping in Fig.\ref{figure3}d reveals that the additional electron pocket shrinks near the $\mathrm{\bar{Y}}$ point, consistent with the edge of the circular feature seen in the Fermi surface of Fig.\ref{figure3}a. In this dispersion, the original TaIrTe$_4$ band is also shifted downward by about $\mathrm{0.1~eV}$, and the tip of the M-shaped horn exhibits a profile similar to that observed upon applying positive gate voltage in Fig.\ref{figure2}c(iii), further confirming that the monolayer TaIrTe$_4$ receives the donated electrons from the Cs atoms. Under this condition, the tip of the M-shaped horn also weakly suggests a ``U"-shape band on top of the valence band, indicating that the conduction band above may start to be filled similar to panel Fig.\ref{figure3}c. Based on the visible $\mathrm{0.1~eV}$ energy shift, we also deduce that the  charge induced by the Cs-dosing is larger than the charge induced by the gating. Thus, we have shown that when the monolayer TaIrTe$_4$ receives additional charges, its  valence bands renormalize before the conduction band starts to fill.

\subsection*{Origin of Cesium adatom induced electron pocket}
\begin{figure*}
\includegraphics[width=0.95\textwidth]{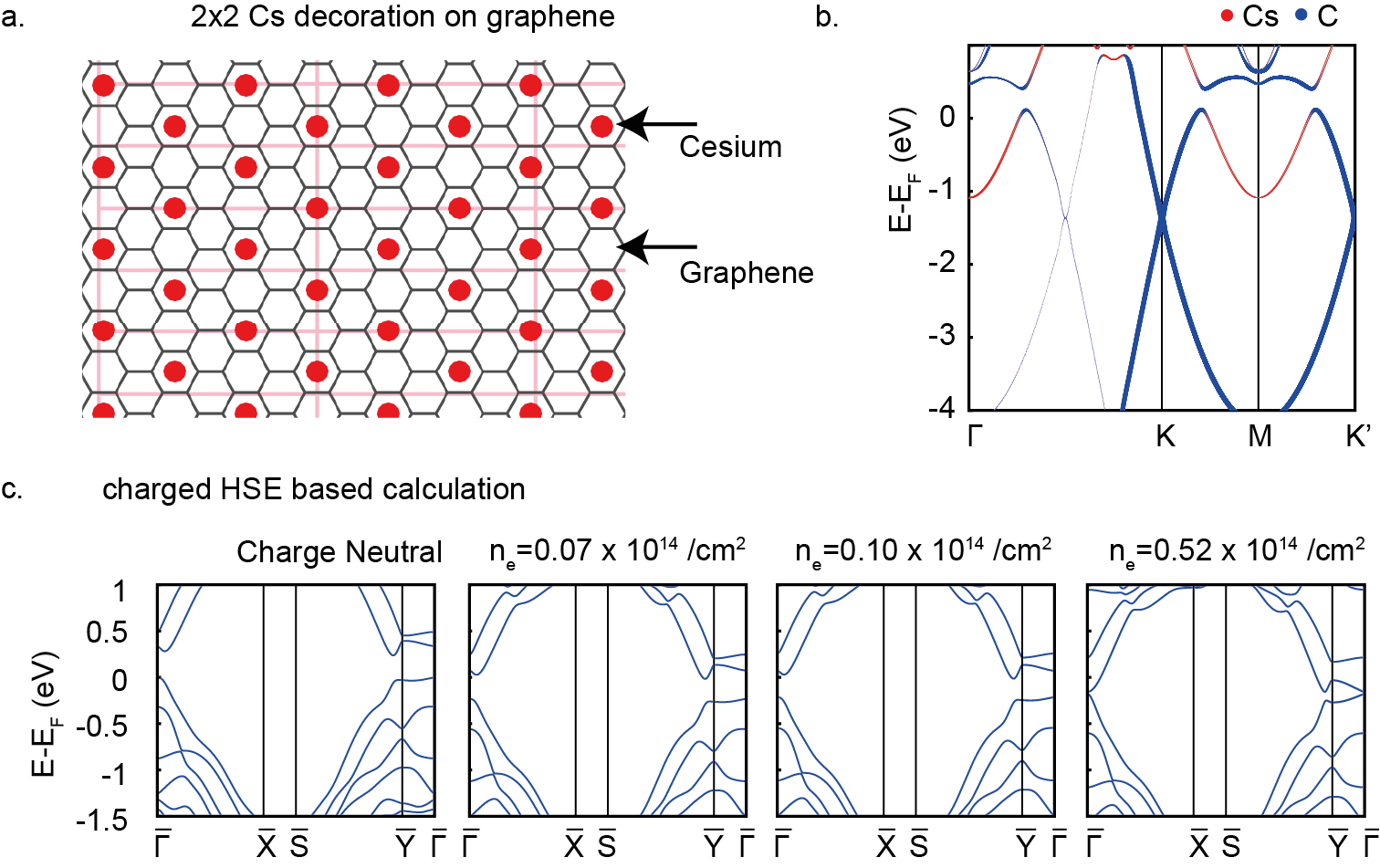}
\caption{\label{figure4}\textbf{Simulated band structure of cesium decorated graphene and charged monolayer TaIrTe$_4$} a. Cesium decorating graphene with a $2\times2$ pattern. The pink rectangle is the unit cell of TaIrTe$_4$ (not included in the simulations), which is not commensurate with the cesium dosed graphene unit cells. b. HSE bandstructure of the cesium decorated graphene (without the TaIrTe$_4$) showing the emergence of cesium-derived electron pockets at the $\Gamma$ and M point, with a significantly smaller spectral weight in the projected band structure. c. HSE band structures of monolayer TaIrTe$_4$ with varying fractional charges showing band gap closing. The corresponding additional electron densities are 0.03,0.05 and 0.25 e/unit cell.}
\end{figure*}

In the following, we perform additional band structure simulations following the ARPES finding and hypothesis above. First, we examine the origin of the electron pocket induced by cesium doping. For this, we can rule out the possibility of alkali metal intercalation between graphene and the TaIrTe$_4$ interface. It is well established that larger Cs adatoms do not intercalate beneath the top graphene layer \cite{watcharinyanon2011rb}, in contrast to smaller alkali atoms such as lithium, which can penetrate the interlayer space \cite{virojanadara2010epitaxial}. This effectively eliminates the likelihood of Cs adatoms residing between the graphene and the monolayer TaIrTe$_4$. Consequently, it is reasonable to attribute the newly observed electron pocket to the graphene monolayer decorated with Cs adatoms,  analogous to the Cs- or Rb-induced electron pockets reported for alkali-decorated graphene\cite{woo2020electron,kaasbjerg2019signatures}. A $2\times2$ or a $\sqrt{3}\times\sqrt{3}$ Cs arrangement on top of graphene would yield  maximum electron densities of $4.8\times10^{14} /\mathrm{cm^{2}}$ and $6.4\times10^{14}/\mathrm{cm^{2}}$, respectively, whereas the observed electron pocket corresponds to about $2\times 10^{14} /\mathrm{cm^{2}}$ and no additional bands appear in the first BZ of pristine graphene. Since the $\sqrt{3}\times\sqrt{3}$ arrangement of atoms on graphene is denser than the $2\times2$, we simulated the band structure of $2\times2$ Cs-decorated graphene, neglecting the TaIrTe$_4$ layer. Fig.\ref{figure4}a shows the proposed Cs-graphene-TaIrTe$_4$ stack, suggesting a vertical superlattice with four times the TaIrTe$_4$ unit cell height, while horizontal periodicity remains unclear due to the incommensurate nature of the stacks. As the TaIrTe$_4$ bands remain essentially unchanged after Cs deposition in the ARPES data (only band renormalization is observed upon receiving additional charges), it is reasonable to neglect the monolayer TaIrTe$_4$ contribution when interpreting the Cs-induced electron pocket.

In Fig.\ref{figure4}b, the band structure of $2\times2$ Cs-decorated graphene is unfolded onto the primitive graphene cell \cite{popescu2012extracting,yang2022electronic}. This procedure maps the supercell eigenstates onto the primitive BZ, enabling direct comparison with the bands of pristine graphene. The line thickness represents the spectral weight of each state (including the structure factor but excluding matrix-element effects), analogous to the intensity distribution observed in ARPES \cite{popescu2012extracting,yang2022electronic}, and the bands are further projected onto elemental contributions. This analysis shows that the Cs atoms generate a new electron pocket at the $\Gamma$ point, as expected. The spectral weight of the electron pocket is lower than that of the Dirac cone at the $K$ point of pristine graphene, consistent with ARPES observations where the new electron pocket intensity is weaker than the graphene $K$-point intensity (see Fig.\ref{S1a}b,c) and comparable to the TaIrTe$_4$ band intensity. The relatively negligible spectral weight of the $2\times2$ supercell K point, located along the $\Gamma \mathrm{K}$ line of the pristine BZ in Fig.\ref{figure4}b, explains the absence of additional unfolded bands within the graphene BZ, aside from the $\Gamma$ point electron pocket. However, the simulation does not account for the unobserved electron pocket at the M point of the graphene (see Fig.\ref{S1a}c). This discrepancy likely arises because the actual arrangement of Cs atoms on top of the graphene lattice is unknown in our sample as they are deposited randomly, and vacancies or other defects may form during Cs deposition. Additionally, the calculation  neglects the potential induced  by TaIrTe$_4$, which could affect the formation of bands beyond the $\Gamma$ point electron pocket. 

\subsection*{Band renormalization from fractional charge calculations}
Next, we further explore the renormalization of the band hypothesis by calculating the HSE band structure with varying  amounts of charge introduced into the system. For a full account of the calculations, refer to Fig.\ref{sfigurecharging} displaying the evolution of the band structure from the removal of 1 electron per unit cell (or adding 1 hole per unit cell) to the addition of 1 electron per unit cell. Fig.\ref{figure4}c shows the key results obtained when adding fractions of an electron into the system. The Fermi level rises immediately to the conduction band even at a fraction of additional electron (see supporting information for Fermi level determination) with the band gap shrinking to zero as more fraction of electron is added, confirming the band renormalization picture upon charging where the band gap closes and the band shape changes. Fig.\ref{sfigurecharging} also shows that the band gap shrinkage and eventual closure also occur when removing one electron per unit cell to the monolayer TaIrTe$_4$. The band gap reduction picture in Fig.\ref{figure4}c at $n_e = 0.52\times 10^{14} / \mathrm{cm}^{2}$ agrees with the gated and dosed ARPES data, which weakly suggest a U-shaped conduction band appearing on top of the valence band. When the band gap is reduced, the valence band renormalization is also visible around the $\mathrm{\bar{Y}}$ point, with the tip of the band becoming sharper like a horn (the M-shaped band), as compared to the neutral condition. As the ARPES data does not show the jump in Fermi level to the conduction band when the band gap is still open and instead it shows a renormalization in Fig.\ref{figure2}b,c and Fig.\ref{figure3}c,d, we reconcile the ARPES data and the DFT calculation by acknowledging the limitation of the stand alone monolayer TaIrTe$_4$ calculation, which ignore the chemical potential arrangement between the graphene and the monolayer TaIrTe$_4$. In other words, the chemical potential between for the two monolayers are pinned to the tip of the monolayer TaIrTe$_4$ valence band until the conduction band appears as the band gap shrinks. From another perspective, it can also be said that any additional electrons received by the monolayer TaIrTe$_4$ which fills the conduction band are depleted towards the graphene. Overall the simulations confirm our interpretation of the ARPES data that the additional charges renormalize the valence band first, prior to filling the unoccupied conduction band and raising the Fermi level.

\section*{Conclusions}
In conclusion, this work has revealed the band structure of a hole-doped, charge neutral, and electron-doped monolayer TaIrTe$_4$, which has been difficult to access due to its air-sensitivity. Encapsulating the monolayer TaIrTe$_4$ with graphene preserves the surface quality for ARPES measurements. Comparison of the valence band observed in ARPES with the DFT calculations using the HSE hybrid functional shows a good agreement. This implies that the monolayer TaIrTe$_4$ has a band gap in the charge neutral condition which is reduced upon both hole and electron doping of the system. Hole doping of the monolayer TaIrTe$_4$ is effective with negatively applied gate voltage, supporting that the graphene and monolayer TaIrTe$_4$ together act as the top metallic plates with hBN acting as the dielectric in this capacitor heterostructure. Meanwhile, applying positive gate voltage and alkali metal deposition, electron dopes the monolayer TaIrTe$_4$, resulting in  renormalization of the monolayer TaIrTe$_4$ band structure before visibly raising its Fermi level in the alkali dosing case. This renormalization of the valence band is well predicted within the HSE framework without considering any strong electron correlation effects apart from the electron exchange interactions intrinsic to the framework. Moreover, the alkali dosing forms a new superstructure with the graphene, which creates a new electron pocket around the $\bar{\Gamma}$ point and significantly increases the Fermi level of the graphene. These findings demonstrate that graphene encapsulated monolayer TaIrTe$_4$ exhibits a non-correlated band structure under electron-doped, charge neutral, and hole-doped conditions, updating the phase diagram proposed in reference \cite{tang2024dual}. Future ARPES measurements on open-surface monolayer TaIrTe$_4$ (or  monolayer TaIrTe$_4$ encapsulated with monolayer hBN)  devices, will be important for fully understanding the  correlated topological phases emerging upon high electron doping as demonstrated by the inability to achieve high charge densities in monolayer TaIrTe$_4$ due to electrostatic screening from graphene layer. In the present study, access to such regimes was limited by electrostatic screening from the graphene layer, which prevented the achievement of comparably high charge-carrier densities to access VHS in conduction band of monolayer TaIrTe$_4$. It will also be of interest to examine how this charge-induced band renormalization depends on strain and on an externally imposed superlattice potential, both of which are expected to modify the topological phase diagram of monolayer TaIrTe$_4$.

\section*{\textbf{Acknowledgments}}
J.K. acknowledges financial support from the U.S. Department Office of Science, Office of Basic Sciences, of the U.S. Department of Energy through award No. DE-SC002549 (postdoc support), NSF-CAREER Award under Grant No. DMR-2339309 and Army Research Office under Grant Number W911NF-26-1-A042, (for graduate student support). This research used resources of the Advanced Light Source, which is a DOE Office of Science User Facility under contract no. DE-AC02-05CH11231. N.M. acknowledges support from the U.S. Department of Energy through grant DE-SC-0019274. This research used computing resources of the National Energy Research Scientific Computing Center (NERSC), a U.S. Department of Energy Office of Science User Facility operated under Contract No. DE-AC02-05CH11231. N.N. acknowledges support from the US Department of Energy, Office of Science, under award number DE-SC0021117 (single-crystal growth and characterization of TaIrTe4).  K.W. and T.T. acknowledge support from the JSPS KAKENHI (Grant Numbers 21H05233 and 23H02052) , the CREST (JPMJCR24A5), JST and World Premier International Research Center Initiative (WPI), MEXT, Japan. We sincerely thank Carlan Gray and Dr. Susan Fullerton in the Department of Chemical and Petroleum Engineering at the University of Pittsburgh for their generous assistance and for providing access to their Atomic Force Microscopy setup.

\section*{\textbf{Contributions}}
J.K. and S.Singh designed and supervised the project. S.S. and R.K.B. fabricated the samples. I.K., A.T., S.S., S.A.E. performed the $\mu$ARPES experiments with the support of C.J., A.B., and E.R. J.T. and Q.M. provided the support for sample and device preparation. S.A.E, A.T., and J.K. analyzed the ARPES data. A.T. performed the AFM measurements. Z.C., C.M., and N.M. performed the DFT calculations. T.Q. and N.N. provided the bulk crystal of tantalum iridium telluride. K.W. and T.T. provided the bulk crystal of hexagonal boron nitride. S.A.E, A.T., Z.C., N.M, and J.K. wrote the manuscript. All authors contributed to the scientific discussion of the results.   

\clearpage
\section*{\textbf{Methods}}

\subsection*{Device fabrication} \label{devicefab}
TaIrTe$_4$ bulk crystals were synthesized using a solution-growth method with Te flux as described in \cite{tang2024dual}. The monolayer TaIrTe$_4$ device was assembled by a dry transfer stacking process in an argon-filled glovebox using a custom-built micromanipulation platform. Bulk TaIrTe$_4$ and graphite crystals were first adhered to low-adhesion blue tape and mechanically cleaved to reduce their thickness. The thinned TaIrTe$_4$ flakes on the tape were then manually picked up with a few-millimeter-thick polydimethylsiloxane (PDMS) stamp and exfoliated onto a plasma-cleaned 300 nm SiO$_2$/Si substrate. Few-layer hBN flakes were prepared similarly by exfoliating bulk hBN onto tape and transferring the selected flakes via PDMS to the substrate. Graphene and graphite were obtained via thermally assisted exfoliation: thinned graphite flakes on tape were transferred to the 300 nm SiO$_2$/Si substrate by heating the tape attached to the substrate at 100-110$^{\circ} \mathrm{C}$ for 8–10 minutes, and cooled for 10-15 minutes before slowly peeling off the tape. 

Layer-by-layer assembly of device was performed via a top-down pickup approach using a composite glass transfer slide consisting of thin film of polycarbonate (PC) on PDMS inside an argon filled glove box. The heterostructure was aligned and transferred into a small gap between pre-patterned Pt(8 nm)/Cr(2 nm) electrodes on a SiO$_2$/Si substrate by melting the PC at 180 $^{\circ} \mathrm{C}$. Prior to the transfer, the electrode region was cleaned in contact-mode Atomic Force Microscopy (AFM) and annealed inside high vacuum at 200°C for 6–8 hours to ensure surface cleanliness. The PC residues were dissolved in anhydrous chloroform for 2 - 3 hours and the stack was cleaned in isopropyl alcohol. In the final device architecture, the bottom graphite gate is connected to one gate electrode, while the top graphene is connected to the opposite gate electrode and serves as both a ground plane and a capping layer to prevent surface degradation of TaIrTe$_4$. The hBN flake between the conductive layers functions as a dielectric. The layer number of the TaIrTe$_4$ flakes was determined using the optical contrast of the exfoliated flakes on 300 nm $SiO_2/ Si$ substrate. The microscope images were white balanced such that the substrate region satisfies R=G=B and appears neutral ensuring the contrast arises from the flake thickness rather than any anomaly in detector sensitivity. The optical contrast increases systematically with layer thickness (inset, Fig.~\ref{sfigure4}a), enabling reliable identification of different thicknesses in the few-layer regime. Due to the air sensitivity of TaIrTe$_4$, additional characterization techniques such as Raman spectroscopy or Atomic Force Microscopy (AFM) could not be performed directly on the monolayer used for ARPES study. Instead, AFM was carried out on separate flakes with comparable optical contrast, and their thicknesses were determined from step-height analysis. A comparison of the AFM topography and height profile of monolayer and bilayer is presented in the Supplementary Information (Fig.\ref{sfigure4}b,c). Given the pronounced difference in optical contrast between monolayer and bilayer, the measured flake can be unambiguously identified as monolayer. AFM measurements were performed in tapping mode using a Bruker Dimension Icon system and $\mu$masch HQ:NSC15/Al BS tip. Topography images were acquired with $512 \times 512$ pixel resolution and analyzed using Gwyddion software  \cite{nevcas2012gwyddion}.

\subsection*{Spatially resolved angle resolved photoemission spectroscopy ($\mu$ARPES)}

Synchrotron based ARPES measurement were performed at Beamline 7.0.2 (MAESTRO) of Advanced Light Source (ALS), Berkeley, CA, USA at a base temperature of approximately 20 K and pressure less than $5 \times 10^{-11}$ torr, using photon energy of $\mathrm{65 - 100~eV}$ with energy and momentum resolution better than 20 meV and 0.01 $\textup{~\AA}^{-1} $ respectively. Before performing ARPES measurements the heterostructure was annealed in ultra high vacuum (UHV) in the range of 180-200 $^{\circ} \mathrm{C}$. The incident X-ray radiation was focused down to spot size of 2 microns and the photoemitted electrons were detected using an R4000 analyser equipped with a deflector. The energy versus momentum slices of the device along the high symmetry direction were obtained by orienting the detector in a preferred direction with respect to the sample. Gating is performed using a Keithley power supply. The cesium getter is available for in-situ dosing while performing ARPES measurement without moving the sample to a different manipulator position.

\subsection*{Density functional theory calculations}

All density functional theory (DFT) calculations were performed using the Vienna Ab Initio Simulation Package (VASP) \cite{1993KresseAbInitioOrigionalPRB} with the projector augmented wave method (PAW) \cite{1994BlochlProjectorAugmentedWaveMethodPRB,1999KresseUltrasoftPseudopotentialsAumentedPRB}. Structural relaxations of monolayer and bilayer TaIrTe$_4$, as well as the cesium-decorated graphene, were carried out using the Perdew-Burke-Ernzerhof (PBE) functional \cite{1996PerdewGeneralizedGradientApproximationPRL}, with van der Waals (vdW) interactions treated by the Tkatchenko-Scheffler (TS) scheme \cite{tkatchenko2009accurate, bucko2013improved, buvcko2014extending}. Electronic structure calculations were performed using the Heyd-Scuseria-Ernzerhof (HSE) \cite{2003JochenHSEFunctionalJourChemPhys, HSEerratum} hybrid functional. PBE results are also reported for comparison. 
The force convergence criterion for structural relaxation was set to $10^{-3}\,\text{eV/\AA}$. The optimized lattice parameters of bulk TaIrTe$_4$, obtained using PBE+TS are: $a = 3.818\,\text{\AA}$, $b = 12.561\,\text{\AA}$, $c = 13.104\,\text{\AA}$. During relaxation of single- and few-layer TaIrTe$_4$ slab models, the in-plane lattice parameters ($a$ and $b$) were fixed to those of the optimized bulk structure. For the relaxation of Cs-decorated graphene, the in-plane lattice parameter was fixed to the equilibrium value of graphene ($a = 2.465\,\text{\AA}$). Cesium doping was modeled by adding one Cs atom per $2\times2$ unit cell, positioned above one of the centers of the carbon hexagons. The relaxed Cs atoms were found to be located at a distance of $3.154\,\text{\AA}$ from the graphene layer. A vacuum spacing of 24\,\AA\; was added along the $c$-axis for the TaIrTe$_4$ and Cs-doped graphene slab models to eliminate interactions between periodic images. A plane-wave cutoff energies of 400\, eV and 520\, eV was used for TaIrTe$_4$ and graphene, respectively. The total energy tolerance for self-consistent field convergence was set to $10^{-7}$\, eV for structural relaxations and $10^{-8}$\, eV for electronic structure calculations. A $\Gamma$-centered mesh with k-spacing of $0.03 \cdot 2\pi\,\text{\AA}{}^{-1}$ was used for relaxations. A denser spacing of $0.01 \cdot 2\pi\,\text{\AA}{}^{-1}$ was employed for electronic structure calculations. Spin-orbit coupling (SOC) was included in all calculations.
For the density of states (DOS) calculations, the tetrahedron method with Blöchl corrections \cite{blochl1994improved} was used to correctly capture the non-differentiability at energy levels where van Hove singularities (vHS) emerge.
To simulate the effect of electron and hole doping, charged calculations were carried out by varying the number of valence electrons within the system relative to the charge-neutral configuration. It should be noted that the Fermi level in DFT is typically determined by setting the chemical potential to conserve the total electron count under a finite-smearing scheme (e.g., Gaussian or Fermi-Dirac) \cite{methfessel1989high}. At very low doping levels, this can keep the reported Fermi level within the gap even when a small conduction-band population is present, due to numerical artifacts from smearing. To avoid this and better reflect the conduction-band filling, we instead reference the Fermi level to the highest occupied state in both Fig.\ref{figure4}c and Fig.\ref{sfigurecharging}.
Additional post-processing, including band-unfolding and visualization, was performed primarily using the vaspvis \cite{dardzinski2022best} and IFermi \cite{ganose2021ifermi} codes.

The assignment of the TaIrTe$_4$ ARPES data to a monolayer, rather than a bilayer or thicker layers, is supported by the DFT results. As shown in Fig.\ref{sfigure3}, increasing the number of layers inevitably introduces additional bands into the band structure, leading to poorer agreement with the experimental data compared to the monolayer case.

\section*{Data availability}
The data that support the findings of this study are available in the Zenodo database \cite{dataset}. Specifically, the underlying ARPES intensity values are embedded within the deposited MATLAB figure files.

\section*{Competing interests}
The authors declare no competing interests.

\clearpage
\bibliography{references}%

\newpage

\section*{Supplementary Information}
\setcounter{figure}{0}
\renewcommand{\thefigure}{S\arabic{figure}}

\begin{figure*}[hbtp]
\includegraphics[width=0.8\textwidth]{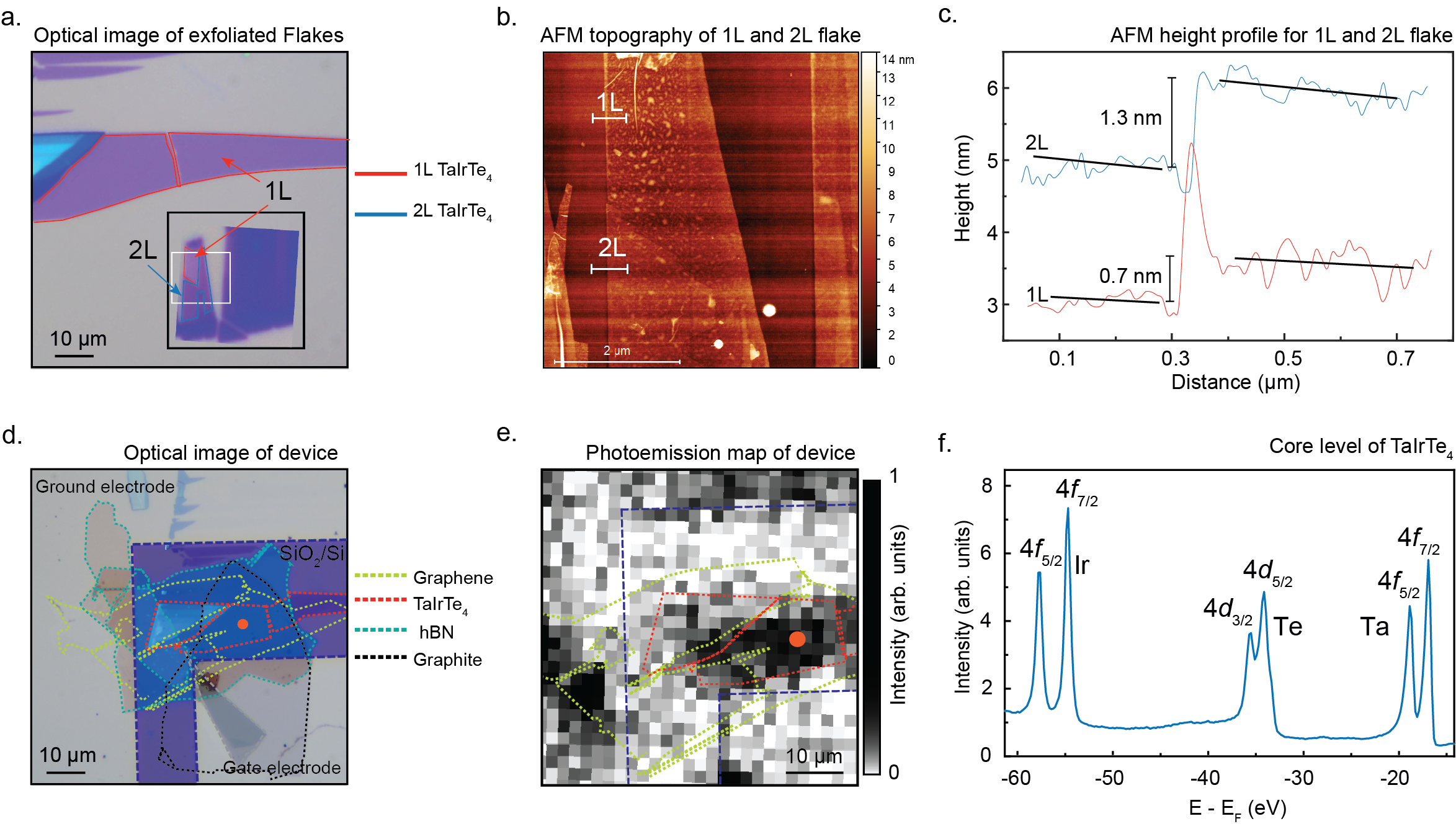}
\caption{\label{sfigure4}\textbf{Optical, AFM and photoemission characterization of monolayer TaIrTe$_4$}  
a. Optical image of exfoliated TaIrTe$_4$ flakes with monolayer (1L, red) and bilayer (2L, blue) regions on SiO$_2$/Si. The white box indicates the area imaged by AFM in (b).
b. AFM topography of region indicated in (a), showing monolayer and bilayer regions. White lines indicate the positions of height profile cuts.
c. AFM height profile for monolayer (red) and bilayer (blue) regions along the line cuts indicated in (b), offset vertically for clarity. Step heights are $0.70 \pm 0.01$ nm and $1.30 \pm 0.01$ nm respectively. The terraces are linearly fitted to the substrate and the flake. The obtained monolayer thickness is consistent with \cite{tang2024dual}.
d. Optical image of the TaIrTe$_4$ device, with different layers of the device highlighted with colored dashed lines. The orange circle marks the spatial location on the device where all the band structure measurements were performed. The top and bottom gray regions correspond to the gate and ground electrodes, respectively, and the inverted L-shaped purple region denotes the SiO$_2$/Si substrate. 
e. Photoemission highlighting TaIrTe$_4$ in the device
f. Core-level photoemission spectrum of the constituent elements in the TaIrTe$_4$ device. 
}
\end{figure*}

\begin{figure*}
\includegraphics[width=0.7\textwidth]{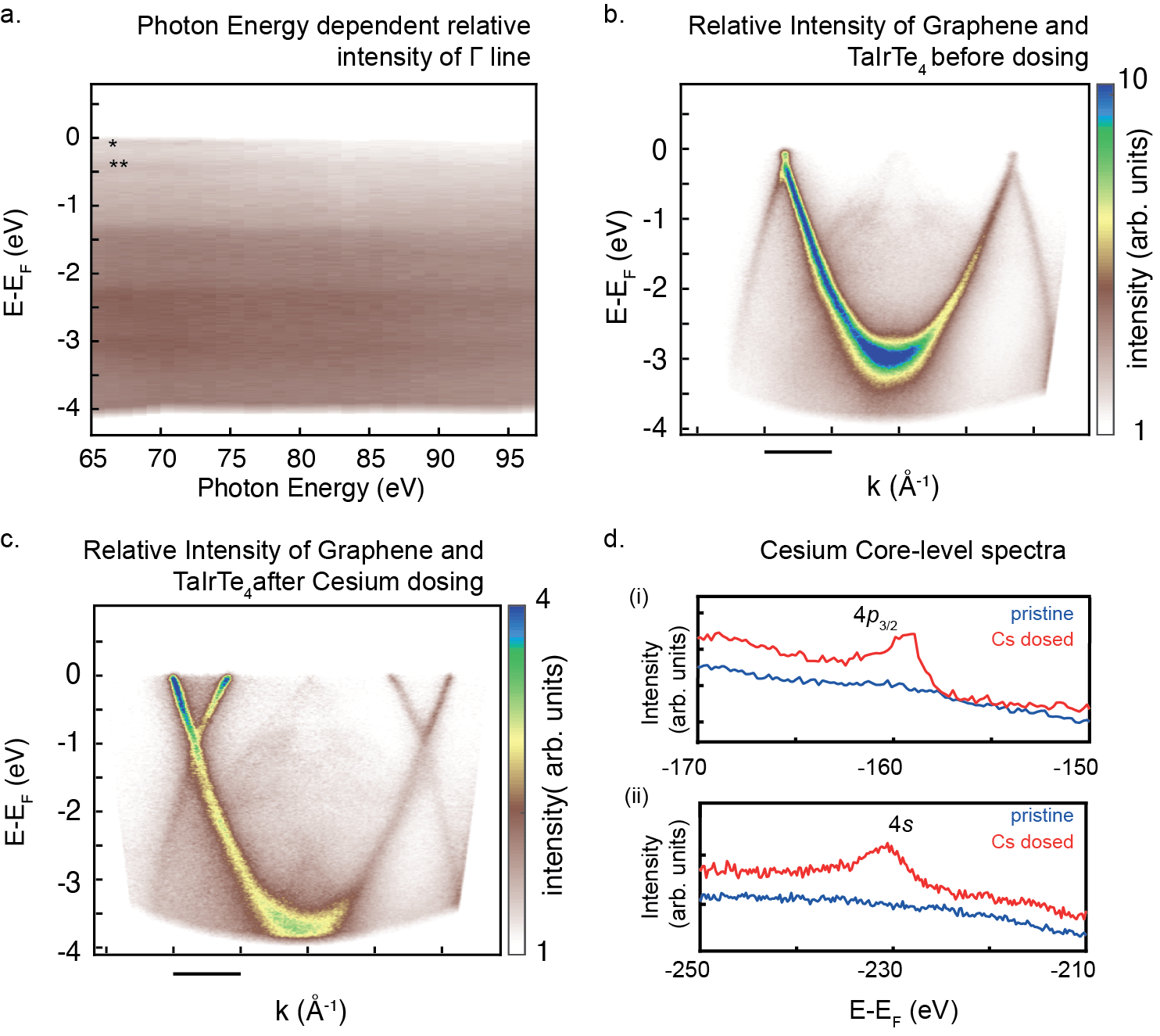}
\caption{\label{S1a}\textbf{Photon energy dependent data and Cesium-dosing supporting data} a. Photon dependent relative intensity of $\bar{\Gamma}$ where photon energy of $\mathrm{70~eV}$ is chosen, showing a relatively flat band as a function of energy indicating 2D property of the bands. * and $\dagger$ bands from the main figure \ref{figure2} are marked b. Dispersion cut of graphene and TaIrTe$_4$ before and c. after the cesium dosing showing that the graphene is significantly brighter than the TaIrTe$_4$ signal. Cut direction is taken from the vertical direction cutting through the KK' point from Figure \ref{figure1}c. Momentum scale-bar is $0.5\,\text{\AA}^{-1}$. d. Cesium 4\textit{p}$_{3/2}$ and 4\textit{s} core level observed on the sample before (blue line) and after dosing (red line).  The intensity of 4\textit{s} peak is relatively higher than that of  4\textit{p}$_{3/2}$.}
\end{figure*}

\begin{figure*}
\includegraphics[width=0.7\textwidth]{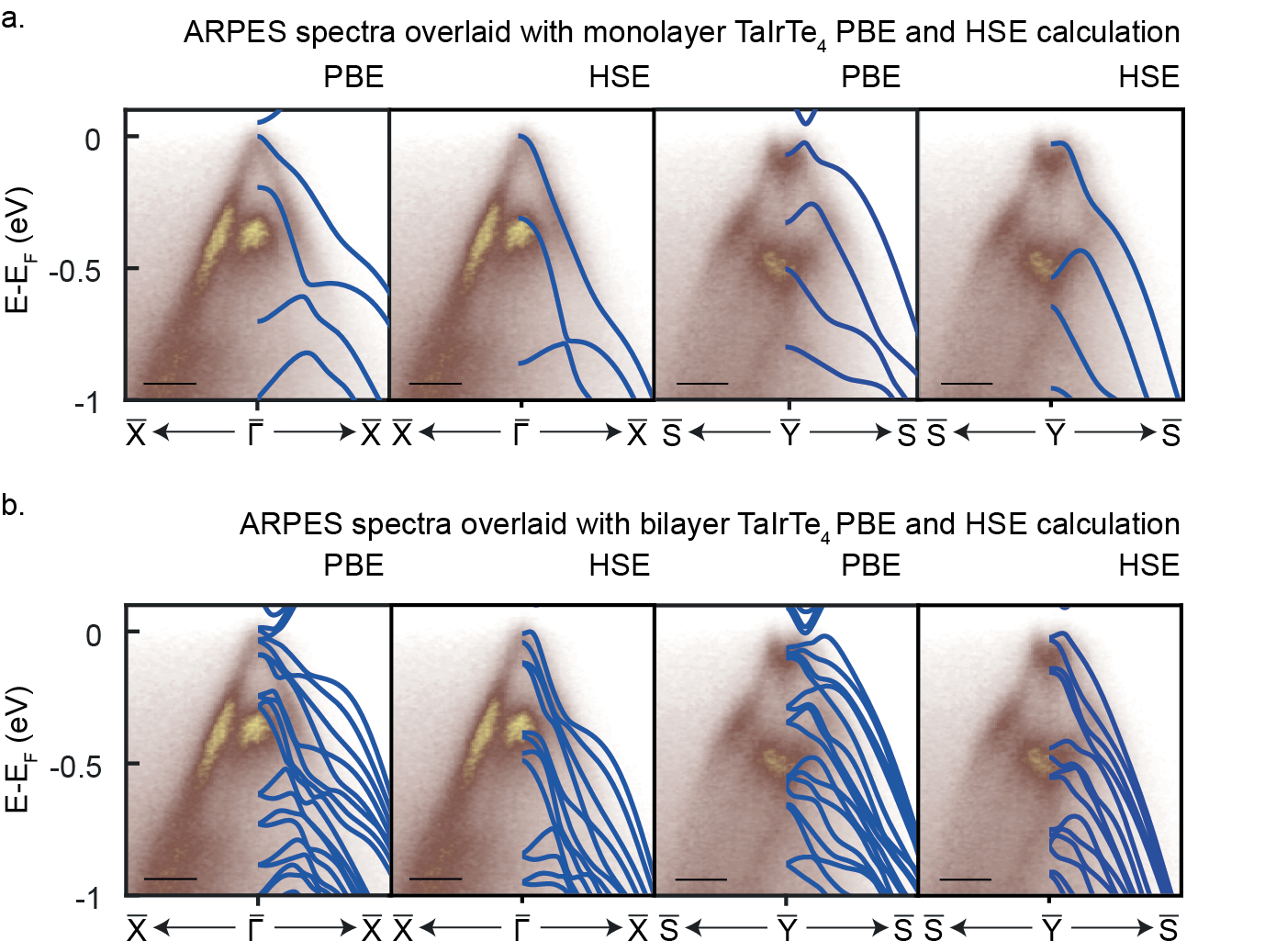}
\caption{\label{S1b}\textbf{DFT band structures compared to ARPES data} a. Comparison to ARPES of bandstructures of a. monolayer TaIrTe$_4$ and b. bilayer TaIrTe$_4$  obtained using the PBE and HSE functionals, showing that the HSE bandstructure of monolayer TaIrTe$_4$ provides the closest agreement with ARPES. The momentum scale-bar is $0.2\,\text{\AA}^{-1}$}
\end{figure*}

\begin{figure*}
\includegraphics[width=0.7\textwidth]{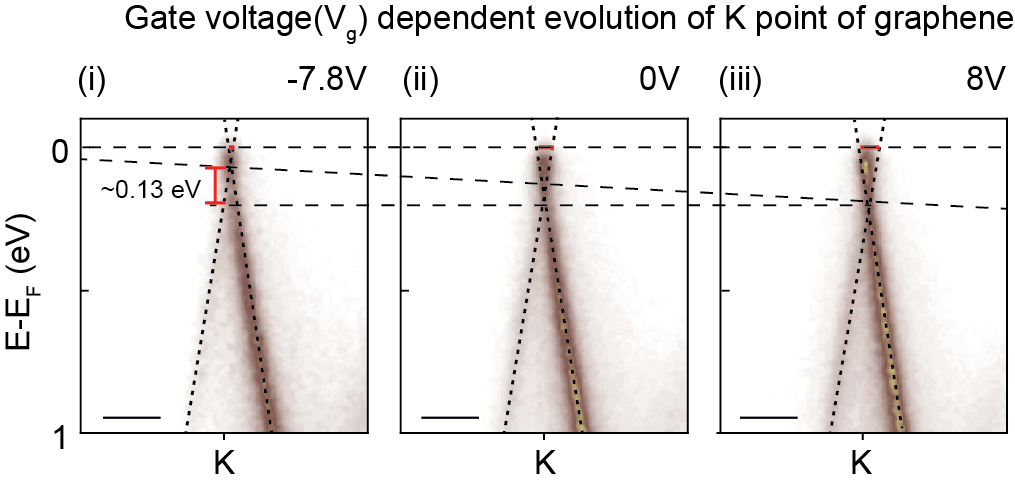}
\caption{\label{S1c}\textbf{Estimation of charge doping on graphene upon electrostatic gating application} a. Graphene cuts obtained from figure \ref{figure2} with estimated linear dispersion cut overlaid to obtain the change in the charge density at the Fermi level of graphene ($\Delta n \approx \pm 1.4\times 10^{12} /\mathrm{cm}^{2}$ for positive and negative bias). Momentum scale-bar is $0.2\,\text{\AA}^{-1}$}
\end{figure*}

\begin{figure*}
\includegraphics[width=0.95\textwidth]
{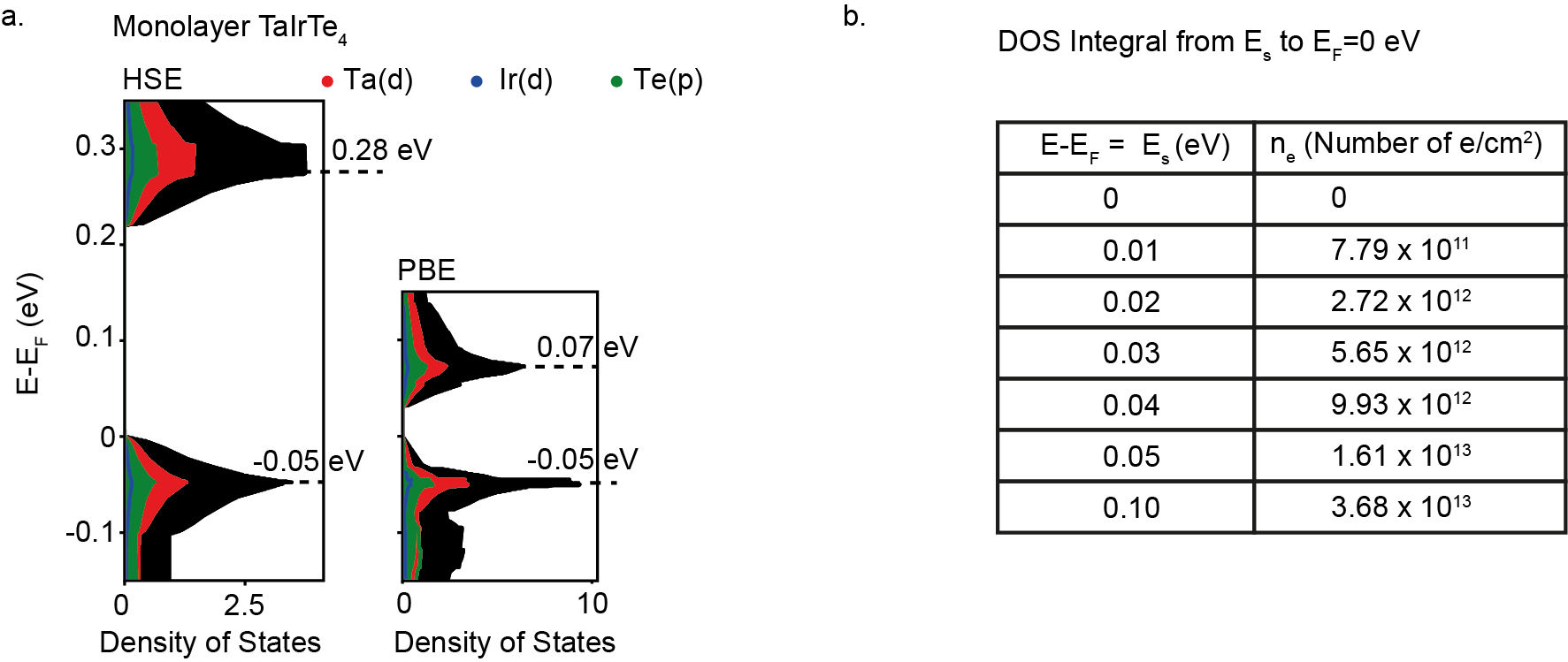}
\caption{\label{sfigure2}\textbf{Density of states (DOS) calculation of monolayer TaIrTe$_4$} a. Density of states obtained using the HSE and PBE functionals. b. Tabulated electron density calculated based on the HSE DOS. The integration was performed from the occupied energy shown in the table up to the Fermi level.}
\end{figure*}

\begin{figure*}
\includegraphics[width=0.5\textwidth]{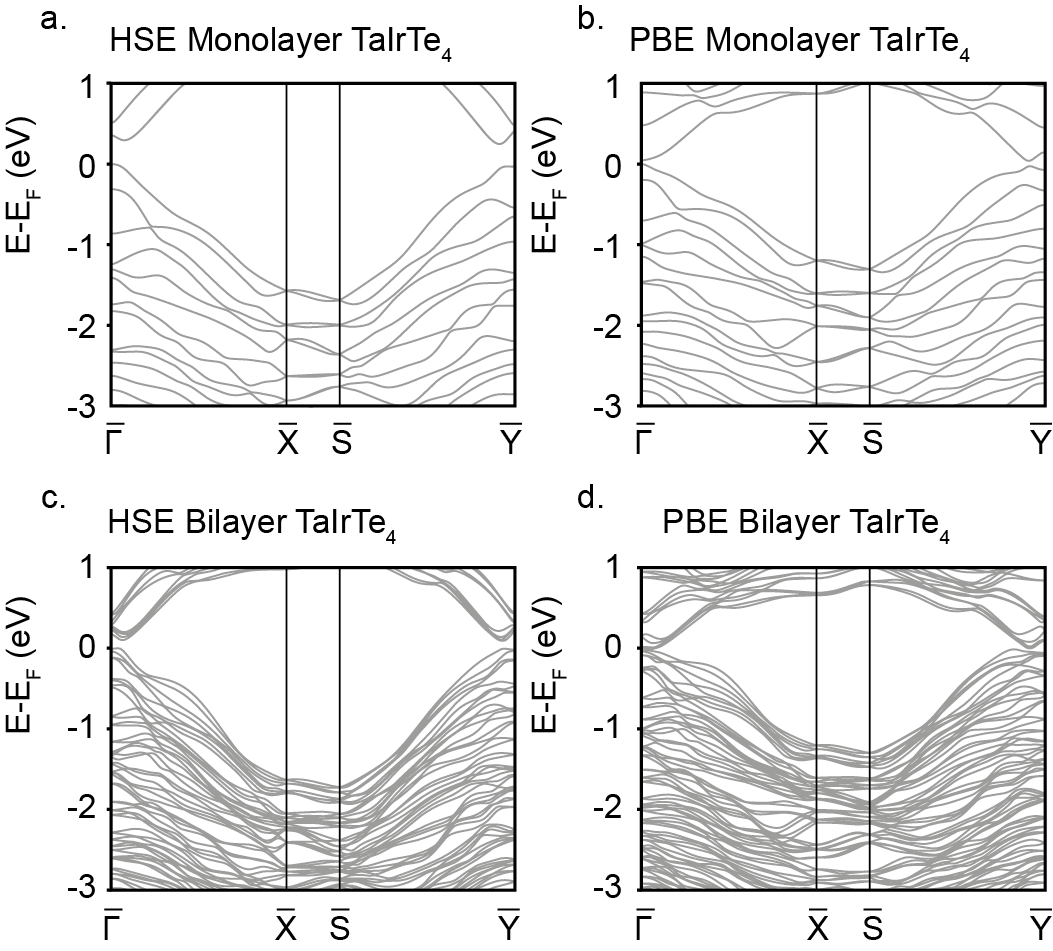}
\caption{\label{sfigure3}\textbf{Comparison of DFT bandstructures of monolayer and bilayer TaIrTe$_4$} Band structure of monolayer TaIrTe$_4$ obtained using a. HSE and b. PBE and band structure of bilayer TaIrTe$_4$ obtained using c. HSE and d. PBE. An expanded energy range is displayed here compared to Figure \ref{S1b}.}
\end{figure*}

\begin{figure*}
\includegraphics[width=1\textwidth]{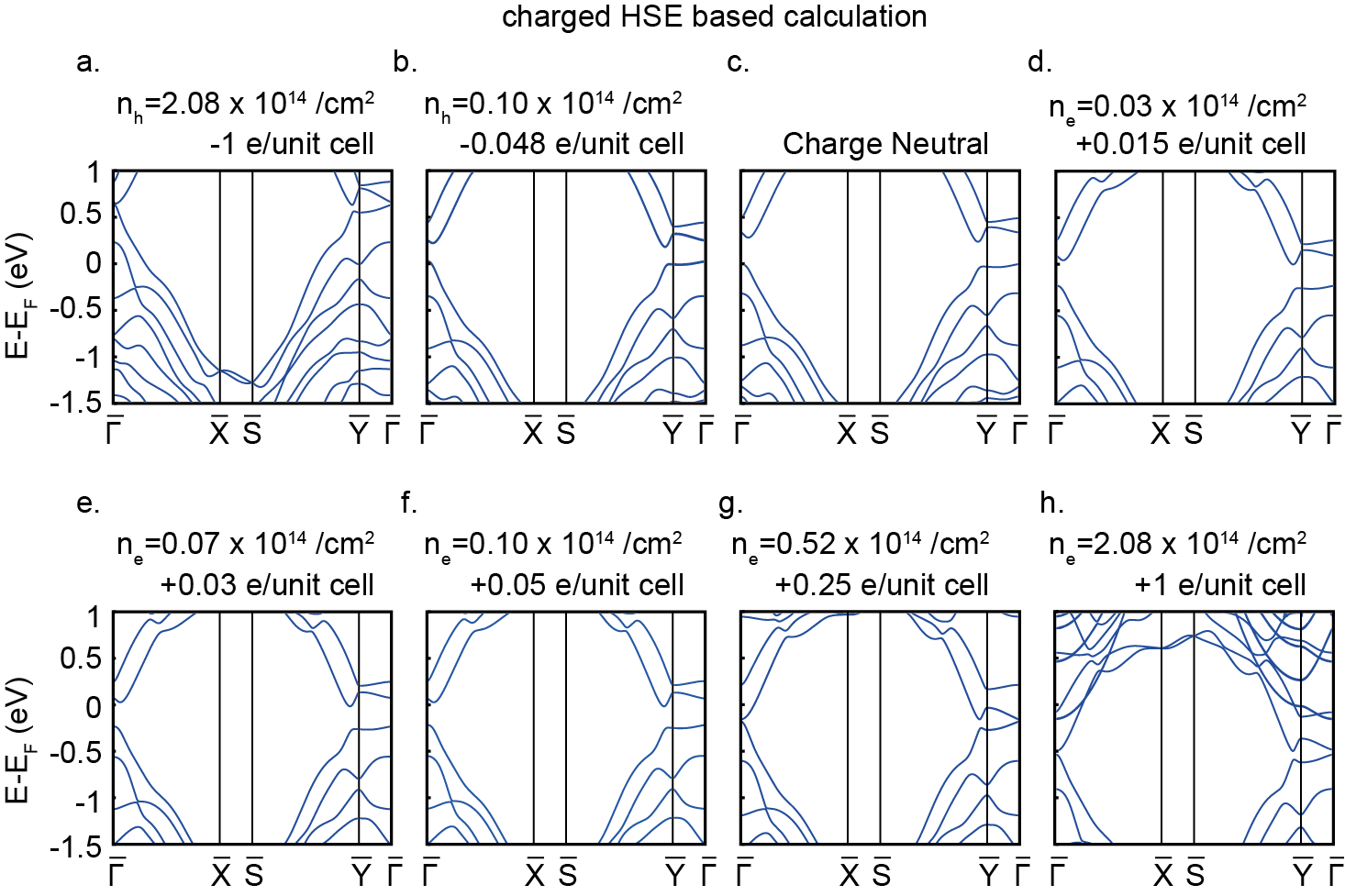}
\caption{\label{sfigurecharging}\textbf{HSE bandstructures of monolayer TaIrTe$_4$ with varying charge.}  Band structures obtained using the HSE functional with a. one electron per unit cell removed, b. 0.048 electrons per unit cell removed, c. charge neutral, d. 0.015 electrons added per unit cell, e. 0.03 electrons added per unit cell, f. 0.05 electrons added per unit cell, g. 0.25 electrons added per unit cell, and h. one electron added per unit cell. The computed band structures reveal the evolution of the band structure upon charging, showing that band renormalization occurs first, followed by eventual band gap closing with increased charging.}
\end{figure*}

\FloatBarrier

\end{document}